# Experiments and Modeling of Mass Transport Phenomena in SiGe Devices

Guangrui (Maggie) Xia
Department of Materials Engineering

The University of British Columbia
Vancouver, Canada
gxia@mail.ubc.ca

*Abstract*—Recent experiments and continuum modeling work on dopant diffusion and segregation, Si-Ge interdiffusion, and defect engineering in SiGe material systems are reviewed. Doping impact on Ge thin film quality and interdiffusion is also discussed. These are relevant to SiGe-based semiconductor devices including SiGe hetero-junction bipolar transistors, metal-oxide-semiconductor field-effect transistors, and Ge-on-Si based photonic devices.

I. INTRODUCTION

In the past three decades, due to the compatibility with Si processing and the capability of mobility, strain and energy bandgap engineering, SiGe, SiGe:C and Ge have been widely used in electronic and optoelectronic devices such as metal oxide semiconductor field effect transistors (MOSFETs) [1], hetero-junction bipolar transistors (HBTs) [2], Ge photodetectors [3], Ge modulators [4], and Ge-on-Si lasers [5].

Although Si and Ge have lots of similar properties, SiGe, SiGe:C and Ge are still different material systems. During the growth and processing, high temperature steps, such as defect annealing, oxidation, deposition, dopant activation annealing are unavoidable. From mass transport point of view, when there is a change in the diffusion media such as a change in Ge and/or carbon (C) concentration or stress or defect concentrations, dopant diffusion and/or segregation changes. Dopants segregate at $Si_{1-x}Ge_x/Si_{1-y}Ge_y$ interfaces or across a graded Ge concentration slope.

Si-Ge interdiffusion should also be considered, as Ge distribution directly impacts dopant diffusion and segregation, bandgap and stress engineering and many electrical and optical properties associated with the $x_{Ge}$. Due to the scaling of semiconductor devices, typical diffusion lengths of Si-Ge interdiffusion after high temperature steps in current technologies are comparable to the thickness of SiGe thin films in the devices, which is in 1 to 100 nm range. Si-Ge interdiffusivity increase exponentially with compressive strain. From the Si end to the Ge end, the interdiffusivity increase by five orders of magnitude with the increase of $x_{Ge}$ in typical thermal annealing temperatures. All these effects can make Si-Ge interdiffusivity comparable or faster than dopant diffusivity.

Defect engineering using C incorporation, oxidation and nitridation can be used to change defect concentrations and engineer dopant profiles. Although those effects have been well studied in Si, oxidation and nitridation have not been well studied or applied in SiGe device fabrication.

These three phenomena, namely dopant diffusion and dopant segregation, Si-Ge interdiffusion, and defect engineering, are important topics for SiGe device structure design and processing, as dopant and Ge distribution is one of the most important factors in determining device characteristics and performance. This abstract will focus on the three mass transport processes in SiGe epitaxial structures. Doping impact on Ge thin film quality and interdiffusion is also discussed.

II. BASE DOPING PROFILE ENGINEEING IN PNP HBTS

Complementary SiGe HBTs have many advantages over an NPN-only technology for numerous analog applications requiring high speed, low noise, and large voltage swing. PNP HBTs use P as the base layer dopant and strained SiGe as the base layer material. Our recent studies addressed three problems related to the mass transport phenomena of PNP HBTs [6], [7].

*A. Effectiveness of C in P profile control*

Although C has been very effective in retarding B diffusion in NPN HBTs, there are limited studies available on the effectiveness of C in PNP SiGe HBTs. Our work quantitatively investigated the C impacts on P diffusion in $Si_{0.82}Ge_{0.18}$:C and Si:C under rapid thermal anneal conditions [6]. The results showed that the C retardation effect on P diffusion is less effective for $Si_{0.82}Ge_{0.18}$:C than for Si:C. In $Si_{0.82}Ge_{0.18}$:C, there is an optimum carbon content at around 0.05% to 0.1%, beyond which more carbon incorporation does not retard P diffusion any more. This can be explained by the decreased interstitial-mediated diffusion fraction $f_I^{P,SiGe}$ to 95% as Ge content increases from 0 to 18%. Empirical models were established to calculate the time-averaged point defect concentrations and effective diffusivities as a function of carbon, and was shown to agree with previous studies on B, P, As and Sb diffusion with C.

## B. Coupled diffusion-segregation of P across graded SiGe

Compared to B, P is harder to control, as it segregates towards lower Ge content layers instead of staying inside SiGe base of higher Ge content. The segregation happens simultaneously with P diffusion. In Ref. [7], experiments were performed with graded SiGe layers for Ge molar fractions ($x_{Ge}$) up to 0.18. A coupled diffusion and segregation model was derived, where the contributions from diffusion and segregation to dopant flux are explicitly shown. The model provides a new approach in segregation coefficient extraction, which is especially helpful for heterostructures with lattice mismatch strains. The diffusion-segregation model for P in SiGe alloys was calibrated and $E_{seg} = 0.5\ eV$ is suggested for the temperature range from 800 to 950 °C.

## C. Thermal nitridation in defect engineering

Thermal nitridation is known to suppress B and P diffusion in Si via exposing bare Si surface in ammonia ambient, where vacancies are injected into Si and retard interstitial diffuser's motion [8]. However, our study showed that this method's effectiveness is limited to low Ge and low C cases [9]. When 0.06% and 0.09% C is present in $Si_{0.82}Ge_{0.18}$, thermal nitridation slightly increases P diffusivity compared to the inert condition.

## III. SI-GE INTERDIFFUSION

Si-Ge interdiffusion happens whenever there is a Ge concentration gradient such as at $Si_{1-x}Ge_x/Si_{1-y}Ge_y$ interfaces and across graded SiGe layers. The interdiffusivity $\widetilde{D}$ is influenced by many factors such as temperature, Ge concentration, stress, doping, and defect density.

## A. Benchmarking model of Si-Ge interdiffusivity

Based on diffusion theories and self-diffusivity data, an interdiffusivity model was established for SiGe interdiffusion under tensile or relaxed strain over $0 \leq x_{Ge} \leq 1$ range for undoped, unstrained or tensile SiGe with low defect densities [10]. It unifies available interdiffusivity data and models over the full Ge range and applies to a wider temperature range. This benchmarking interdiffusivity model and/or relevant experiment data have been implemented in major process simulation tools, and the calculation results showed good agreement with experimental and literature data under furnace annealing and soak and spike rapid thermal annealing conditions [10].

## B. Strain and oxidation impact

The impact of strain and strain relaxation on Si-Ge interdiffusion in epitaxial SiGe heterostructures was systematically investigated in [11-14]. $x_{Ge}$ studied ranged from 0.3 to 0.75, and the temperature range was 720–880 °C. Tensile strain up to 1% in $Si_{0.7}Ge_{0.3}$ layer has no observable impact on the interdiffusion [11], while compressive strain was shown to increase Si-Ge interdiffusivity exponentially [12, 13]. Theoretical analysis showed that strain field can add on to the interdiffusion driving force from the concentration gradient. This effect can be included in the apparent interdiffusivity term $\widetilde{D}_{apparent}$ [13]. Thermal oxidation had little impact on interdiffusion in relaxed $Si_{1-x}Ge_x$/compressive $Si_{1-y}Ge_y/Si_{1-x}Ge_x$, in which x and y are in the range of 0.15 to 0.58 [15].

## C. Doping impact on Ge film quality and interdiffusion

For SiGe with low Ge molar fractions for HBT applications, C, P and B were all shown to enhance Si-Ge interdiffusion [16]. On the high Ge end, Ref. [17-18] investigated P doping effect on interdiffusion and the modeling of the effect. Ge/$Si_{1-x}Ge_x$/Ge multi-layer structures with $0.75 < x_{Ge} < 1$, a mid-$10^{18}$ to low-$10^{19}$ cm$^{-3}$ P doping and a dislocation density of $10^8$ to $10^9$ cm$^{-2}$ range were studied. The medium P-doped sample shows an accelerated Si-Ge interdiffusivity, which is 2–8 times of that in the undoped sample. The doping dependence of the Si-Ge interdiffusion was modelled by a Fermi-enhancement factor. The results indicate that the interdiffusion in high Ge fraction range with n-type doping is dominated by $V^{2-}$ defects.

After this study, Ge/Si structures with three different dopants (P, As and B) and those without intentional doping were investigated [19]. All samples have a smooth surface (roughness < 1.5 nm), and the Ge films are almost entirely relaxed. Etch pit density (EPD) measurements showed that B doped Ge films have EPD above $1 \times 10^8$ cm$^{-2}$, while P and As doping can reduce the EPD to be less than $10^6$ cm$^{-2}$ without annealing. Cautions need to be taken in interpreting the EPD results in n-Ge, as the preferable etching of threading dislocations didn't work as good as in the undoped Ge, which cause a significant underestimation of threading dislocation density [20].

However, photoluminescence (PL) study shows that unannealed P and As-doped Ge samples have much higher PL intensities than those annealed samples due to much less interdiffusion [21]. The PL intensities from the unannealed n-doped Ge are also comparable to that of a benchmarking sample with 4 times of doping [21]. This offers a new method to fabricate high-quality Ge-on-Si films without defect annealing procedure, which avoids undesired interdiffusion associated with defect annealing. A quantitative model of Si-Ge interdiffusion under extrinsic conditions across the full Ge range was established including the dislocation-mediated diffusion. The Kirkendall effect has been observed. The results are of technical significance for the structure, doping, and process design of Ge-on-Si based devices, especially for photonic applications.

IV. ACKNOWLEDGEMENTS

Texas Instruments (TI), Crosslight Software Inc. and National Science and Engineering Research Council of Canada (NSERC) are acknowledged for funding the studies in Ref. [6-7, 9-10, 13-14, 17-19]. Drs. Simon Li, Zhiqiang (Leo) Li, Yue (Fred) Fu from Crosslight, Mr. Hiroshi Yasuda, Dr. Stanley Philips, Dr. Manfred Schiekofer and Mr. Bernhard Benna from TI are acknowledged for the research collaborations in these studies.